\documentclass[aps, 10pt, prb, showkeys, twocolumn, superscriptaddress]{revtex4-2}
\usepackage{graphicx, xcolor, amssymb, amsmath, amsfonts, enumitem}
\usepackage[utf8]{inputenc}
\usepackage[normalem]{ulem}
\usepackage[colorlinks=true, linkcolor=blue, citecolor=blue, urlcolor=blue]{hyperref}
\usepackage{url, soul, ulem, textgreek, bm, comment}
\usepackage[dvipsnames]{xcolor}
\usepackage{makecell}
\usepackage{appendix}
\usepackage{orcidlink} 

\newcommand{\apctpadd}{Asia Pacific Center for Theoretical Physics, Pohang, Gyeongbuk, 37673, Republic of Korea}
\newcommand{\postechadd}{Department of Physics, Pohang University of Science and Technology, Pohang, Gyeongbuk 37673, Republic of Korea}

\begin{document}

\author{Sanghoon Lee\orcidlink{0009-0001-2248-5534}}
\affiliation{\apctpadd}

\author{Kyoung-Min Kim\orcidlink{0000-0003-2468-3152}}
\email{kyoungmin.kim@apctp.org}
\affiliation{\apctpadd}
\affiliation{\postechadd}

\title{Mobility-edge-embedded Hofstadter butterfly from a tilt-induced quasiperiodic potential}


\keywords{Hofstadter butterfly, Quasiperiodicity, Mobility edge, Localization, Fractal dimension, Square lattice}

\begin{abstract}
The Hofstadter butterfly (HB) and mobility edges (MEs) are hallmark phenomena of quasiperiodic systems, yet their interplay remains elusive.
Here, we demonstrate their coexistence within a tilt-induced quasiperiodic potential on a square lattice, giving rise to a “mobility-edge-embedded Hofstadter butterfly” (MEE-HB).
This potential is generated by aligning a periodic potential at an angle relative to the lattice axes---a configuration readily accessible in optical lattice experiments.
Using a tight-binding model, we show that the MEE-HB manifests as a fractal energy splitting pattern hosting MEs that separate extended and localized states.
Our Harper-like equation shows that the fractal pattern originates from one-dimensional quasiperiodic potentials, while MEs stem from effective long-range hopping. 
Notably, the MEE-HB exhibits a fractal dimension of \(0.8\)--\(1.0\), significantly exceeding the \(0.4\)--\(0.6\) range of the standard butterfly, indicating a denser spectrum.
Our findings establish tilt-induced potentials as a versatile platform for exploring the interplay between fractal structures and localization.
\end{abstract}

\date{\today}

\maketitle

\section{Introduction}

Quasiperiodic systems provide a fascinating platform for exploring exotic quantum phenomena that are absent in their periodic counterparts \cite{SOKOLOFF1985189, Macia_2006}.
A fundamental example is the Hofstadter butterfly (HB) \cite{PhysRevB.14.2239}, whose self-similar fractal sub-gap structure represents a paradigmatic quantum fractal in condensed matter physics.
Another prominent example is the mobility edge (ME) \cite{PhysRevLett.61.2144}, which defines the energy-dependent separation between coexisting extended and localized states.
Unlike random disordered systems \cite{doi:10.1126/science.1209019}, quasiperiodic lattices can accommodate MEs in dimensions below three, and their deterministic nature bypasses the requirement for randomness, providing a highly controllable environment for studying localization transitions.
These two phenomena have spearheaded theoretical research for decades, revealing a wide array of underlying mechanisms and variants \cite{kraus2013four, lellouch2014localization, du2018floquet, yao2019critical, PhysRevLett.86.1062, liu2022anomalous, sun2024nonhermitian, lee2026structural, 7std-nbqw, PhysRevLett.114.146601}.
Furthermore, recent experimental advancements have enabled the realization of HBs \cite{PhysRevLett.109.106402, PhysRevLett.111.185301} and the observation of MEs \cite{luschen2018single, An2021InteractionsMobilityEdges, PhysRevLett.120.160404, PhysRevLett.126.040603} in optical lattices, as well as HBs in moiré superlattices of electronic systems \cite{Dean2013, Ponomarenko2013, doi:10.1126/science.1237240, Nuckolls2025}.
Given these developments, a natural question arises: can a single quasiperiodic system accommodate these two distinct phenomena simultaneously?
Despite the naturalness of this inquiry, their coexistence has been largely unexplored, and the interplay between HBs and MEs remains elusive.

\begin{figure}[t]
    \centering
    \includegraphics[width=\linewidth]{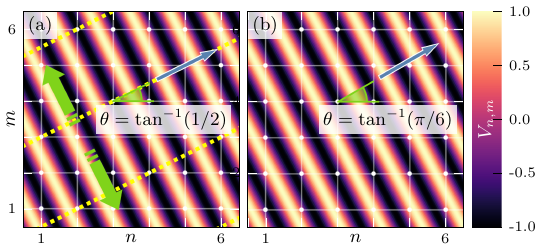}
    \caption{
    \textbf{Tilt-induced quasiperiodic potentials.}
    (a) Onsite potential \(V_{n,m} = \lambda \cos[2\pi(n\cos\theta + m\sin\theta)]\) for a rational slope (\(\tan\theta = 1/2\)), where the modulation vector (blue arrow) is tilted from the horizontal axis by an angle $\theta$ (green arc).
    The potential renders the system quasiperiodic along the modulation direction while remaining periodic in the transverse direction (green arrows), thereby forming a supercell of repeating stripes (separated by dashed yellow lines).
    (b) Same potential for the irrational slope (\(\tan\theta = \pi/6\)).
    In this case, the underlying translational symmetry of the square lattice is fully broken.
    White dots denote square lattice sites, and the potential strength is set to \(\lambda=1\).
    }
    \label{fig:schema}
\end{figure} 

In this work, we demonstrate the coalescence of HBs and MEs---a phenomenon we term the ``mobility-edge-embedded Hofstadter butterfly'' (MEE-HB). As a concrete platform, we propose a two-dimensional (2D) system of non-interacting particles subjected to a one-dimensional (1D) sinusoidal potential aligned at an angle relative to the principal axes of a square lattice [Fig.~\ref{fig:schema}]. This configuration is readily accessible using ultracold atoms in optical lattices. By solving the corresponding tight-binding model, we demonstrate the emergence of the MEE-HB, whose energy spectra exhibit fractal band splitting in the energy-tilt angle space that embeds MEs (Sec.~\ref{sec:2D_results})---a feature absent in the standard Harper-Hofstadter model \cite{PhysRevB.14.2239}. We attribute the formation of the MEE-HB to the effective 1D quasiperiodicity and long-range hopping terms arising from the tilt-induced modulation, as described by our 1D Harper-like equation (Sec.~\ref{sec:reduction_to_1d}). Furthermore, box-counting analysis \cite{LIEBOVITCH1989386} reveals that the fractal dimension of the MEE-HB spectrum (0.8--1.0) significantly exceeds that of the standard HB (\(0.4\)--\(0.6\)), serving as a hallmark characteristic of the system (Sec.~\ref{sec:fractal_dimension}). Finally, we discuss experimental realizations in optical lattices (Sec.~\ref{sec:experimental_setup}) and highlight the advantages of this tilt-induced approach over alternative configurations (Sec.~\ref{sec:other_possible_routes}).

\section{Tilt-dependent localization} \label{sec:localization}

We consider a tight-binding Hamiltonian on a square lattice subjected to a 1D onsite potential, whose orientation relative to the lattice is controlled by a tilt angle \(\theta\). 
The Hamiltonian in second-quantized form reads as follows:
\begin{equation}
\begin{aligned}
\label{eq:H}
	H = &\; t \sum_{n,m} \left( \hat{c}_{n+1,m}^{\dagger} \, \hat{c}_{n,m} + \hat{c}_{n,m+1}^{\dagger} \, \hat{c}_{n,m} + \mathrm{h.c.} \right) \\
      &+ \sum_{n,m}  V_{n,m} \, \hat{c}_{n,m}^{\dagger} \, \hat{c}_{n,m}, 
\end{aligned}
\end{equation}
where \(\hat{c}_{n,m}^{\dagger}\) (\(\hat{c}_{n,m}\)) creates (annihilates) a particle at lattice site \((n,m)\), and \(t\) denotes the nearest-neighbor hopping amplitude. 
Throughout this work, we set \(t=1\).
The onsite potential is given by
\begin{equation}\label{eq:potential}
	V_{n,m} = \lambda \cos\left[ 2\pi ( n\cos\theta + m\sin\theta ) \right],
\end{equation}
where \(\lambda\) is the potential strength, and the vector \((\cos\theta,\sin\theta)\) defines the direction of the onsite potential. It is sufficient to consider $\theta \in [0, \pi/4]$, as the Hamiltonian is invariant under $\theta \rightarrow \theta + \pi/2$ and $\theta \rightarrow \pi/2 - \theta$ (up to lattice rotations or reflections). In the limiting case $\theta = 0$, the potential $V_{n,m}$ reduces to a uniform value $\lambda$, and the Hamiltonian simplifies to a trivial nearest-neighbor hopping model. For $\theta \in (0, \pi/4]$, however, the phase $2\pi(n\cos\theta + m\sin\theta)$ induces a quasiperiodic modulation that breaks the underlying translational symmetry. We therefore refer to this system as a ``tilt-induced quasiperiodic model (TQM).''

\begin{figure}[t]
    \centering
    \includegraphics[width=\columnwidth]{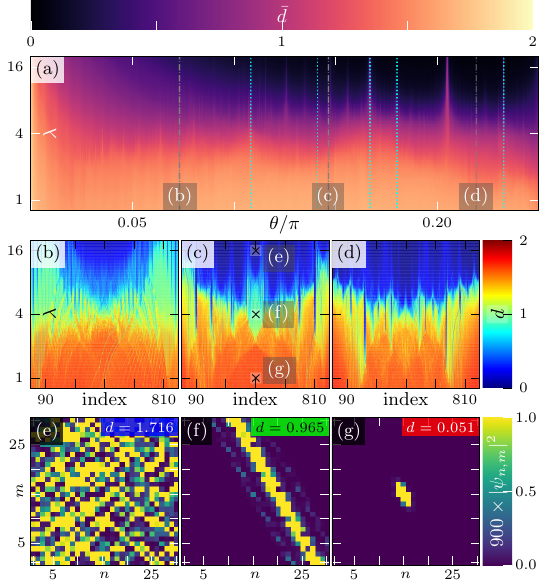}
    \caption{
    \textbf{Tilt-dependent localization.} 
    (a) Spectrum-averaged participation dimension \(\bar{d} = -{\ln\langle\mathrm{IPR}\rangle}/{\ln L}\) in the \((\theta/\pi, \lambda)\) plane.
    Cyan dotted lines mark the angles satisfying \(\sin\theta=1/2\), \(1/3\), \(2/3\), \(3/7\), and \(15/28\).
    (b)--(d) Eigenstate-resolved participation dimension at \(\theta = 0.073\pi\), \(0.146\pi\), and \(0.219\pi\), respectively, as marked by the vertical dashed lines in (a).
    The values \(d\simeq2\), \(1\), and \(0\) correspond to extended, line-like, and point-like localized character, respectively.
    The horizontal axis labels eigenstates ordered by increasing energy.
    (e)--(g) Real-space density profiles for the states marked by crosses in (c), illustrating extended states and localized states with line-like and point-like profiles, respectively.
    In all panels, the \(\lambda\)-axis is plotted on a logarithmic scale.
    The results are obtained by solving the tilt-induced quasiperiodic model (TQM) on a \(30\times 30\) square lattice with open boundary conditions (OBCs). 
    }
    \label{fig:avg_ipr_2d}
\end{figure}

The corresponding single-particle Schr\"odinger equation for the wavefunction amplitude \(\psi_{n,m}\) reads
\begin{equation}
\begin{aligned}\label{eq:schrodinger_eqn}
	E\psi_{n,m} = & \;\psi_{n+1,m} + \psi_{n-1,m} + \psi_{n,m+1} + \psi_{n,m-1} \\[3pt]
	&+ \lambda \cos\left[2\pi(n\cos\theta + m\sin\theta)\right]\psi_{n,m}.
\end{aligned}
\end{equation}
To characterize the localization properties of the system, we numerically diagonalize the Hamiltonian on a finite \(L\times L\) square lattice with open boundary conditions (OBCs).
Then we compute the inverse participation ratio (IPR) for all eigenstates which serves as a diagnostic of spatial localization.
For a given normalized eigenstate \(\psi\), the IPR is defined as
\begin{equation}\label{eq:IPR_2D}
	\mathrm{IPR} = \sum_{n,m} |\psi_{n,m}|^{4}.
\end{equation}
In a two-dimensional system, the magnitude of the IPR can be compared with the characteristic finite-size scales expected for different spatial supports: 
\(\mathrm{IPR}\sim L^{-2}\) for a state spread over the full \(L\times L\) area, \(\mathrm{IPR}\sim L^{-1}\) for a state occupying an \(O(L)\) line-like region, and \(\mathrm{IPR}\sim L^{0}\) for a state confined to an \(O(1)\) region.
To differentiate these three cases, we define the participation dimension for a finite width \(L\)
\begin{gather}\label{eq:deff}
    d = -{\ln \mathrm{IPR}}/{\ln L},
\end{gather}
so that \(d\simeq 2\), \(1\), and \(0\) indicate extended, line-like, and point-like localized profiles of states, respectively.

As an initial overview, Fig.~\ref{fig:avg_ipr_2d}(a) shows
\begin{gather}
    \bar{d} = -{\ln\langle\mathrm{IPR}\rangle}/{\ln L},
\end{gather}
where \(\langle \text{IPR} \rangle\) denotes an averaged IPR over all eigenstates at fixed \((\theta,\lambda)\).
For \(\lambda<4\), \(\bar{d}\) remains close to two over most of the tilt-angle range, indicating that the spectrum is dominated by extended states.
Around \(\lambda\simeq 4\), \(\bar{d}\) develops an angular dependence.
It remains close to two at small tilt angles, but approaches values near unity as \(\theta\) increases.
As \(\lambda\) increases further, the region with \(\bar{d} < 1\) extends toward small tilt angles, signaling that a larger fraction of the spectrum is occupied by localized states with line-like or point-like profiles.

Aside from this global trend, \(\bar{d}\) exhibits pronounced enhancements along vertical cuts defined by \(\sin\theta = b/a\) (for \(b/a \leq \sqrt{2}/2\)) or \(\cos\theta = b/a\) (for \(b/a \geq \sqrt{2}/2\)), where \(a\) and \(b\) are positive coprime integers, as illustrated by the cyan dotted lines in Fig.~\ref{fig:avg_ipr_2d}(a).
Along these vertical cuts, the potential becomes periodic in either the \(n\) or \(m\) direction, restoring translational symmetry along one of the lattice axes.
Consequently, at these angles which are one-axis-commensurate along a single lattice direction, bulk states remain extended along the corresponding periodic direction according to Bloch's theorem.
Interestingly, the average IPR also vanishes at \(\theta = \pi/4\), although the translational symmetry is fully broken in all lattice directions.
Our findings reveal that the bulk states extend along the diagonal direction of the system.
In both cases, a small fraction of states remains localized at specific boundary sites due to the OBCs used~\cite{Pradhan_2016, Li24}.
Although OBCs can be replaced by a PBCs along the periodic direction at the one-axis commensurate angles, we use OBCs for generic values of \(\theta\).
Detailed analysis of the one-axis-commensurate and diagonal cases can be found in the Supplementary Note 1.



Figures~\ref{fig:avg_ipr_2d}(b)--(d) show how the dimensional character of the eigenstates changes across the spectrum. 
Here we plot the participation dimension \(d\) for each eigenstate as a function of the energy-sorted state index and the potential strength \(\lambda\) for three representative tilt angles, \(\theta = 0.073\pi\), \(0.146\pi\), and \(0.219\pi\), respectively.
In all three cases, states with \(d \lesssim 1\) appear clearly once \(\lambda\) becomes larger than approximately \(\lambda = 4\).
However, the distribution of these regions in the spectrum depends strongly on the tilt angle.
For the smallest tilt angle [Fig.~\ref{fig:avg_ipr_2d}(b)], point-like regions with \(d\simeq 0\) first appear near the center of the spectrum and in narrow windows near the spectral edges, while many states between them show the line-like value \(d\simeq 1\).
As the tilt angle increases [Figs.~\ref{fig:avg_ipr_2d}(c) and (d)], the \(d \simeq 0\) region progressively expands and penetrates deeper toward the spectral edges, while the \(d\simeq1\) states remain mostly near the boundary between the \(d\simeq 2\) and \(d\simeq 0\) regions.
These results indicate the emergence of MEs.
The sharp contrast in the participation dimension \(d\) shows that extended, line-like, and point-like localized states occupy different parts of the spectrum.
The real-space profiles in Figs.~\ref{fig:avg_ipr_2d}(e)--(g) make this dimensional distinction explicit.
They show representative states selected from the center of the spectrum in Fig.~\ref{fig:avg_ipr_2d}(c) at \(\lambda=1\), \(4\), and \(16\), respectively.
As \(\lambda\) increases, the state changes from an extended state to a line-localized state and then to a point-localized state.


\section{Mobility-edge-embedded butterfly spectra}
\label{sec:2D_results}

\begin{figure*}
    \centering
    \includegraphics[width=\textwidth]{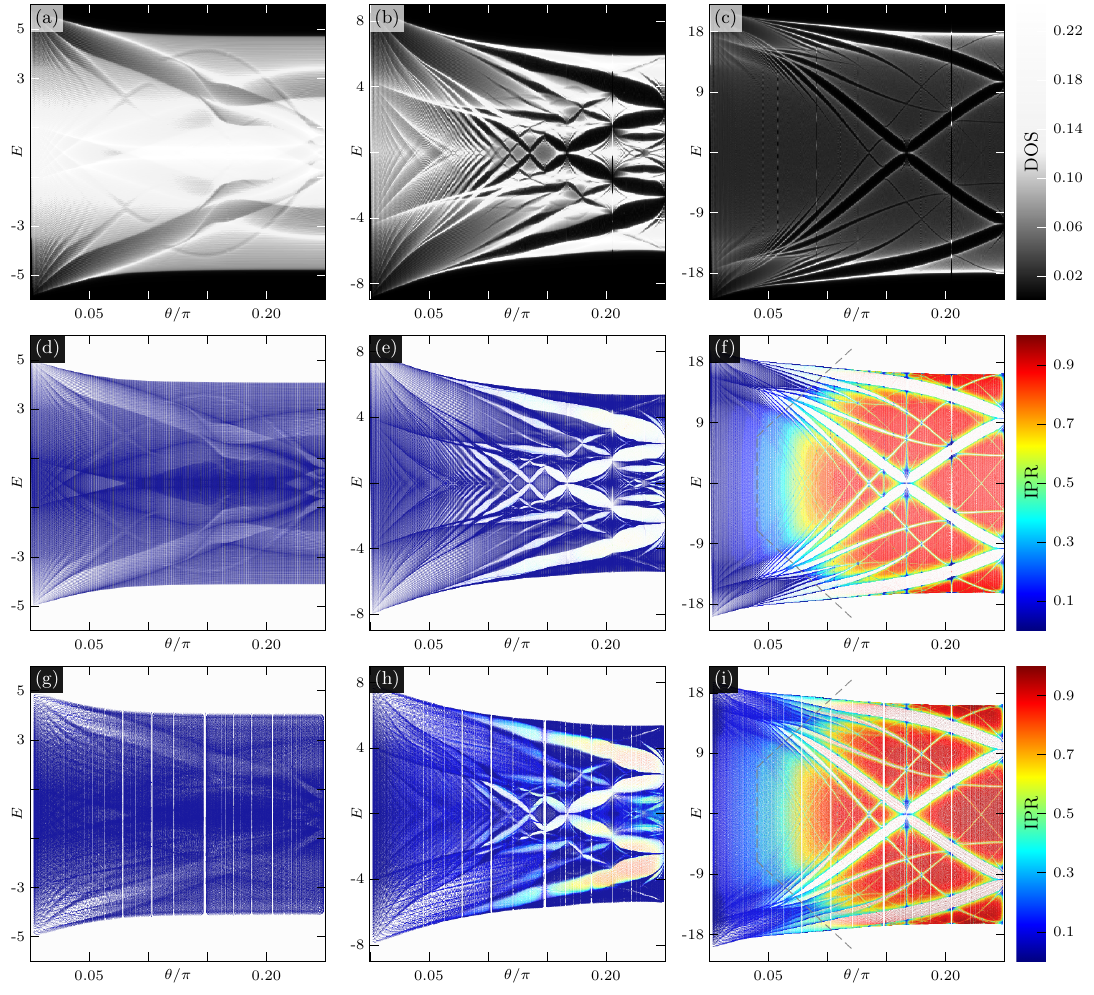}
    \caption{
    \textbf{Hofstadter butterflies (HBs) of the tilt-induced quasiperiodic model (TQM).} 
    (a)--(c) Density of states (DOS) in the $(\theta, E)$ plane for (a) $\lambda=1$, (b) $\lambda=4$, and (c) $\lambda=16$. 
    In (a), the DOS reveals a pseudo-gap forming wing-shaped patterns.
    In contrast, the spectra in (b) and (c) exhibit self-similar, recursive subband structures with full bandgaps.
    (d)--(f) IPR maps in the $(\theta, E)$ plane corresponding to (a)--(c). 
    In (d) and (e), IPR values remain uniformly low across the bulk states. However, in (f), the IPR values bifurcate into extended (blue) and localized (red) regions. 
    The dashed lines, defined by $\text{IPR} = 0.3$, mark the boundary between these two regions and represent the mobility edges (MEs) of the spectra.
    The combination of self-similar subband structures and extended bulk states identifies (b) and (e) as a conventional HB, while the presence of both self-similar subbands and MEs in (c) and (f) identifies the system as a mobility-edge-embedded Hofstadter butterfly (MEE-HB). 
    (g)--(i) IPR maps obtained from the effective Harper-like model [Eq.~\eqref{eq:effective_1d}] for the same values of $\lambda$ as in (d)--(f). 
    The effective model precisely reproduces both the energy spectra and IPR distributions observed in the TQM. 
    The tilt angle is defined by $\tan \theta = p/q$, where $(p, q)$ are coprime integers.
    }
    \label{fig:all_HB}
\end{figure*}

To investigate the fractal energy splitting arising from the potential in Eq.~\eqref{eq:potential},
we calculate the energy spectrum for three representative potential strengths ($\lambda = 1, 4,$ and $16$) over the tilt-angle range $\theta \in [0, \pi/4]$.
We visualize the spectra in the \((\theta, E)\) plane through a density of states (DOS) obtained by Lorentzian broadening of the discrete eigenvalues, with the spectral weight normalized by the total number of states.
Figure~\ref{fig:all_HB}(a) illustrates the evolution of the resulting DOS map for a weak potential strength ($\lambda = 1$) in the $(\theta, E)$ plane.
At the limiting case of $\theta = 0$, the DOS exhibits a continuous distribution across the entire energy range $E \in [-3, 5]$, with a pronounced concentration at the band center ($E = 1$).
This behavior stems from the restoration of the system's periodic nature at $\theta = 0$, where the onsite potential reduces to a constant energy offset of $\lambda = 1$. 
For $\theta>0$, the onset of quasiperiodicity causes this continuum to break down into finer structures characterized by localized high-density peaks in the low-density background.
As $\theta$ increases, these peaks branch self-similarly, yielding a hierarchy of wing-shaped regions in the $(\theta, E)$ plane.
These regions retain a small but finite DOS, forming pseudo-gaps (PGs) rather than true subband gaps.
This feature distinguishes the observed recursive pattern from a conventional HB-type fractal, as it lacks a fully opened gap structure.
This spectral evolution with a PG structure serves as a precursor to the well-defined HB hierarchy emerging at higher potential strengths.
Furthermore, the IPR distribution remains uniformly low across the entire spectrum [Fig.~\ref{fig:all_HB}(d)], confirming that all eigenstates in this PG regime are predominantly extended.

At an intermediate potential strength ($\lambda=4$), the DOS develops multiple wing-shaped bandgap regions in the $(\theta, E)$ plane, establishing a well-defined subband structure [Fig.~\ref{fig:all_HB}(b)].
Within this structure, larger wings recursively embed smaller, identically shaped features, creating a self-similar pattern that spans multiple scales. Consequently, the spectrum manifests the signature fractal splitting characteristic of a standard HB.
Furthermore, the bulk states, which correspond to the high-density profiles in the DOS map, exhibit minimal IPR values, indicating their extended nature [Fig.~\ref{fig:all_HB}(e)].
In contrast, a small fraction of in-gap states, appearing as sparsely scattered points within the subband gaps, are localized at the system's boundaries due to the OBCs used~\cite{Pradhan_2016, Li24}.
Despite the presence of these localized states, the overall physical behavior in the large-system limit is governed by the bulk states, as evidenced by their dominant fraction in Fig.~\ref{fig:all_HB}(b).
Therefore, these two defining features---the self-similar recursive subband structure and the prevalence of extended bulk states---identify the energy spectra of the system as a standard HB-type fractal.

As the potential strength $\lambda$ increases, the widths of the subband gaps in the recursive structure gradually narrow relative to the total band width, as illustrated for $\lambda = 16$ in Fig.~\ref{fig:all_HB}(c).
Furthermore, the recursive subband structure now accommodates a coexistence of both extended and localized states within the spectrum [Fig.~\ref{fig:all_HB}(f)]. Specifically, at small angles ($\theta \lesssim 0.04\pi$), all states remain extended. Localized states first emerge once $\theta$ exceeds approximately $0.04\pi$, while states near the spectral edges maintain their extended nature. As $\theta$ increases, the localized region spreads toward the spectral edges; for $\theta \gtrsim 0.13\pi$, the majority of the spectrum becomes dominated by localized states. These extended and localized regions are largely separated by white dashed lines in Fig.~\ref{fig:all_HB}(f), which represent the MEs of the spectra. The emergence of these MEs within the self-similar structure identifies the energy spectra of the system as an unconventional HB-type fractal, namely the MEE-HB proposed in this work.
It is worth noting that the delocalization of bulk states persists at commensurate angles and at $\theta = \pi/4$, even within the localized regime ($\theta \gtrsim 0.13\pi$), as discussed in the preceding section.
We find that the subband gap structure first becomes prominent at $\lambda \approx 2$. Furthermore, MEs first emerge within this subband structure at $\lambda \approx 6$ in the high-tilt-angle regime. 
This suggests that $\lambda \approx 2$ constitutes the threshold for the transition from PGs to HBs, while $\lambda \approx 6$ marks the onset of the transition from standard HBs to MEE-HBs. For a more detailed discussion, refer to Supplementary Note 2.

\section{Effective Harper-like equation} \label{sec:reduction_to_1d}

\begin{figure}[t]
    \centering
    \includegraphics[width=\columnwidth]{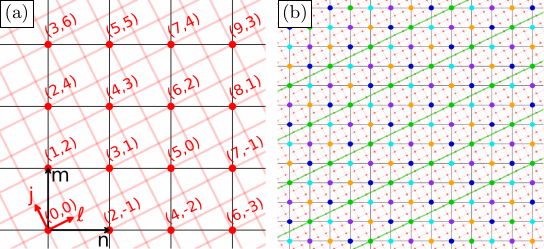}
    \caption{
    \textbf{Rotated coordinates for the Harper-like model.}
    (a) Black lines display the original square lattice, while red lines represent the rotated coordinate system, with red markers denoting the points where the two lattices overlap. These coincident points are the only ones maintained during the mapping of the TQM into the effective Harper-like model. The numbers signify the rotated integer coordinates $\ell=2n+m$ and $j=-n+2m$, where $(n,m)$ are the original lattice coordinates. A rational slope of $\tan\theta = 1/2$ is used here.
    (b) A wide-field view of the pristine and rotated lattices shown in (a). Different sublattice sites within each supercell strip are classified according to $j \pmod 5$ and indicated by distinct colors. The green lines correspond to a fixed residue class, explicitly demonstrating that the lattice repeats periodically along the $j$-direction under the translation $j \to j+5$. 
    }    
    \label{fig:coordinate_rotation}
\end{figure}

To reveal the physical mechanism underlying the emergence of MEE-HB, we derive an effective 1D Harper-like model for the TQM defined in Eq.~\eqref{eq:schrodinger_eqn} for rational slopes \(\tan\theta = p/q\) where \(\gcd(p,q) = 1\).
Using \(\cos\theta = \alpha q\) and \(\sin\theta = \alpha p\), where \(\alpha = 1/\sqrt{p^{2} + q^{2}}\), the potential becomes quasiperiodic \(V_{n,m} = \lambda \cos [2\pi \alpha (qn + pm)]\), which is effectively one-dimensional with respect to the combined index \(\ell \equiv qn + pm\).
To reorganize the hopping terms, we introduce a complementary integer coordinate transverse to the direction of the quasiperiodic potential \(j \equiv -pn+qm\), so that the lattice sites are relabeled by
\begin{gather}\label{eq:transformation}
	\begin{bmatrix}  \ell     \\ j         \end{bmatrix}
	=
	\begin{bmatrix}  q  &  p  \\ -p  &  q  \end{bmatrix}
	\begin{bmatrix}  n        \\  m        \end{bmatrix}.
\end{gather}
Under a unit lattice shift, \((n,m)\to(n \pm1 , m)\) and \((n,m)\to(n,m\pm1)\), the new coordinates transform as \((\ell,j)\to(\ell \pm q, j\mp p)\) and \((\ell,j)\to(\ell \pm p, j \pm q)\), respectively.
The resulting coordinate transformations are illustrated in Fig.~\ref{fig:coordinate_rotation}(a).
To distinguish the transformed representation from the original lattice wavefunction \(\psi_{n,m}\), we denote the wave amplitude on the \((\ell,j)\) lattice by \(\chi_{\ell,j}\).
Then Eq.~\eqref{eq:schrodinger_eqn} is rewritten in the \((\ell, j)\) coordinate as
\begin{align}\label{eq:schrodinger_eqn_lj}
	E\chi_{\ell,j} =
    &\;\lambda\cos(2\pi\alpha\ell)\chi_{\ell,j} \\[3pt]
	   &+ \chi_{\ell+q,\,j-p} + \chi_{\ell-q,\,j+p} + \chi_{\ell+p,\,j+q} + \chi_{\ell-p,\,j-q}, \notag
\end{align}
where the two directions play distinct roles: the quasiperiodic potential varies only along \(\ell\), while \(j\) labels the direction perpendicular to the modulation.
The transverse direction repeats after a finite translation, $(\ell, j) \to (\ell, j+D)$ with $D=p^{2}+q^{2}$, resulting in a superlattice structure [Fig.~\ref{fig:coordinate_rotation}(b)].
Consequently, PBCs along the $j$-direction can be imposed by formally regarding two distant supercells as identical.
The tilted geometry therefore induces an effective superlattice perpendicular to the modulation, so that the eigenstates can be labeled by a transverse Bloch momentum \(\nu\).
We therefore write \(\chi_{\ell,j}\) in the Bloch form:
\begin{gather}\label{eq:bloch_j}
    \chi_{\ell,j} = \exp(i\nu j)\,\varphi_{\ell}^{(\nu)}.
\end{gather}
For a finite transverse superlattice of length \(N_{j}D\) with periodic boundary conditions, \(\nu\) takes the discrete values \(\nu=2\pi n / (N_jD)\), where \(n\) runs from \(0\) to \(N_jD-1\).
Formally, the Bloch ansatz diagonalizes the transverse translation structure of Eq.~\eqref{eq:schrodinger_eqn_lj}, and we obtain
\begin{align}\label{eq:effective_1d}
    E\varphi_{\ell}^{(\nu)}
    =   &\;\lambda\cos(2\pi\alpha\ell)\,\varphi_{\ell}^{(\nu)} \\[3pt]
        & + \exp\left(-i\nu p\right) \varphi_{\ell+q}^{(\nu)}
          + \exp\left(+i\nu p\right) \varphi_{\ell-q}^{(\nu)} \notag \\[3pt]
        & + \exp\left(+i\nu q\right) \varphi_{\ell+p}^{(\nu)}
          + \exp\left(-i\nu q\right) \varphi_{\ell-p}^{(\nu)}.\notag 
\end{align}
For each transverse Bloch momentum \(\nu\), this representation yields a 1D Harper-like equation characterized by a quasiperiodic onsite potential and finite-range hopping.
It nevertheless differs from the conventional Harper model in two key aspects. First, it involves multi-range hopping terms rather than the standard nearest-neighbor hopping.
Second, both the hopping distances and the incommensurate frequency $\alpha$ are determined by the coprime integer pair $(p, q)$.
Consequently, the spectral evolution of the TQM arises from the simultaneous modification of both the potential and hopping terms, as varying $(p, q)$ modifies both components of the effective 1D chain.

Figures~\ref{fig:all_HB}(g)--(i) show the corresponding angle-resolved spectra for three representative potential strengths, \(\lambda = 1\), \(4\), and \(16\).
The effective one-dimensional model precisely reproduces the same spectral evolution seen in Figs.~\ref{fig:all_HB}(d)--(f): the PG regime at \(\lambda=1\), a HB at \(\lambda=4\), and the MEE-HB at \(\lambda=16\).
The quasiperiodic cosine potential, which varies depending on the choice of \((p,q)\) pairs, provides the Harper-type mechanism that generates fractal energy splitting, whereas the long-range hopping distances \(p\) and \(q\) are known to promote mobility edges in a quasiperiodic chain~\cite{biddle2010predicted, biddle2009localizationA, biddle2011localizationB}.
Therefore, the MEE-HB arises from the interplay between a Harper-like quasiperiodic potential and beyond-nearest-neighbor hopping in the reduced chain given in Eq.~\eqref{eq:effective_1d}.

It is worth noting that the effective Harper-like model, when employing the PBC in the \(j\) direction for rational slopes \(\theta_{\mathrm{rat}} = \tan^{-1}(p/q)\), displays only extended and line-like localized states.
However, the model can be extended to genuine irrational slopes by expanding the phase of the potential, \(\Phi = 2\pi(\cos\theta n + \sin\theta m)\), around the rational tilt angle \(\theta_{\mathrm{rat}}\).
Specifically, the phase is approximated as \(\Phi \simeq 2\pi\alpha l + 2\pi\alpha \delta\theta \, j\), where \(\delta\theta = \theta - \theta_{\mathrm{rat}}\) represents the small deviation of the irrational tilt angle \(\theta\) from a given rational tilt angle \(\theta_{\mathrm{rat}}\).
With this substitution, the potential in Eq.~\eqref{eq:schrodinger_eqn_lj} becomes \(\lambda \cos(2\pi\alpha l + 2\pi\alpha \delta\theta \, j)\), which is quasiperiodic in both the \(l\) and \(j\) directions.
Consequently, the system loses its periodicity in the \(j\) direction, enabling the effective model to describe the emergence of point-like localized states alongside extended and line-like states.
This result is consistent with the comprehensive numerical findings obtained from the TQM shown in Fig.~\ref{fig:all_HB}(f).

The isospectral dual representation of Eq.~\eqref{eq:effective_1d} along the \(j\)-direction provides additional insights into the emergence of the MEE-HB in the TQM.
Applying the Aubry--Andr\'e Fourier transform~\cite{aubry1980analyticity} to Eq.~\eqref{eq:effective_1d} yields its isospectral dual model~\cite{lee2026structural}:
\begin{align}\label{eq:isodual}
    E\tilde{\varphi}_{k}^{(\nu)}
    =&\; \frac{\lambda}{2}\left(\tilde{\varphi}_{k+1}^{(\nu)}+\tilde{\varphi}_{k-1}^{(\nu)}\right) \\[3pt]
    &+ 2\!\left[
    \cos\!\left(\nu p-2\pi\alpha q k\right)
    +\cos\!\left(\nu q+2\pi\alpha p k\right)
    \right]\tilde{\varphi}_{k}^{(\nu)}.
    \notag
\end{align}
where $\tilde{\varphi}_{k}^{(\nu)}$ denotes the dual amplitude at site \(k\),
\begin{gather}
    \varphi_{\ell}^{(\nu)} = \sum_{k\in\mathbb{Z}}\tilde{\varphi}_{k}^{(\nu)}\exp\left(i(2\pi\alpha k)\ell\right).
\end{gather}
In the dual representation, the finite-range hoppings of Eq.~\eqref{eq:effective_1d} are mapped onto a quasiperiodic onsite term composed of two cosine contributions, while the original monochromatic onsite potential becomes nearest-neighbor hopping with amplitude \(\lambda/2\).
Therefore, for fixed \(\nu\), Eq.~\eqref{eq:isodual} takes the form of a nearest-neighbor chain subject to a structured quasiperiodic potential, making its relation to the Harper family more transparent.
Crucially, the cosine potentials govern the formation of the self-recursive subband hierarchy and also break the simple self-dual structure of the standard Aubry--Andr\'e model.
Such broken self-duality is a known route to MEs in one-dimensional quasiperiodic systems with longer-range hopping or multi-harmonic quasiperiodic potentials~\cite{soukoulis1982localization,li2017mobility,li2020mobility,lee2026structural}.
The structural versatility of the TQM---specifically its mapping onto various equivalent representations---suggests that the MEE-HB is not tied to a particular formulation, but may arise more broadly in physical systems sharing the same underlying mathematical structure.

\section{Fractal dimension} \label{sec:fractal_dimension}

\begin{figure}[t]
    \centering
    \includegraphics[width=\columnwidth]{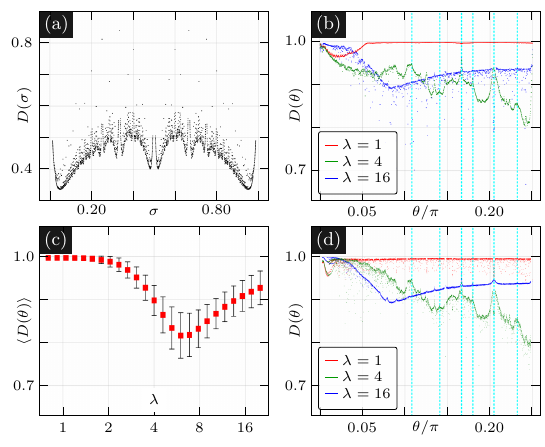}
    \caption{
    \textbf{Fractal dimension from box-counting analysis.}
    (a) Fractal dimension $D(\sigma)$ of the Harper model calculated at rational frequencies $\sigma=p/q$ under periodic boundary conditions (PBCs).
    (b)--(c) Fractal dimension \(D(\theta)\) of the TQM.
    In (c), \(\langle D(\theta)\rangle\) denotes the average of \(D(\theta)\) over \(\theta\in[0,\pi/4]\), and the error bars represent the standard deviation.
    (d) $D(\theta)$ calculated from the isospectral dual model for rational slopes $\tan\theta = p/q$.
    The effective model accurately reproduces the fractal-dimension profile of the original TQM shown in (b), except for the small-angle regime ($\theta < 0.05\pi$) where the original results exhibit artifacts due to finite-size effects.
    Cyan dotted lines mark the angles satisfying \(\sin\theta=1/2\), \(1/3\), \(2/3\), \(3/7\), \(15/28\), and \(\cos\theta=4/5\).
    }
    \label{fig:fractal_dimension}
\end{figure}

To characterize the HBs arising from the TQM, we perform a box-counting analysis~\cite{vicsek1992fractal, falconer2014fractal, 8txw-m2cp} on their energy spectra.
In this approach, the energy range of a given spectrum is partitioned into bins (boxes) of size $\epsilon$.
We then determine the number of occupied boxes $N(\epsilon)$, defined as those containing at least one energy level.
The box-counting dimension $D$, which we preferentially refer to as the ``fractal dimension,'' is defined as:
\begin{equation}
    D \equiv -\lim_{\epsilon\to 0}\frac{\ln N(\epsilon)}{\ln \epsilon}.
\end{equation}
Practically, we extract $D$ by performing a linear fit of $\ln N(\epsilon)$ against $\ln \epsilon$ for sufficiently small values of $\epsilon$.
Further technical details of this regression procedure are provided in the Methods section.
The resulting $D$ serves as a robust metric for the self-similar distribution of energy levels.
A value of $D=1$ corresponds to a continuous spectrum without subband formation.
Conversely, $D < 1$ signifies a sparse spectrum characterized by the emergence of subband gaps, with $D$ representing the degree of such spectral fragmentation.
As a benchmark, we first calculate the fractal dimension of the standard Harper model~\cite{hofstadter1976energy}:
\begin{equation}\label{eq:harper}
    E g_n = g_{n+1} + g_{n-1} + 2\lambda\cos(2\pi\sigma n - \nu) g_n,
\end{equation} 
where $g_n$ denotes the wave amplitude at site $n$. 
Figure~\ref{fig:fractal_dimension}(a) displays the fractal dimension $D(\sigma)$ for this model.
Over a wide range of $\sigma$, the values of $D(\sigma)$ remain significantly below unity, predominantly ranging between $0.3$ and $0.6$.
These values provide a quantitative measure of substantial spectral fragmentation, a feature clearly reflected in the sparse energy spectrum of the canonical HB [Fig.~\ref{fig:other_butterfly}(d)].
Notably, \(D(\sigma)\) approaches unity as \(\sigma\) tends to \(0\) or \(1\).
In these limits, the onsite potential in Eq.~\eqref{eq:harper} becomes constant, so the model reduces to a free tight-binding chain up to an overall energy shift.
The gaps between subbands disappear, and the spectrum becomes continuous.
The mean value of $D(\sigma)$, averaged over $\sigma$, is 0.54, which closely aligns with the results reported in previous work \cite{wilkinson1994spectral}.

We next evaluate the fractal dimension $D(\theta)$ of the HBs using the box-counting method to quantify the degree of spectral fragmentation. For this analysis, the energy spectra are preprocessed to exclude in-gap, boundary-localized states, allowing us to capture the intrinsic spectral fragmentation of the self-similar bulk energy levels while eliminating finite-size artifacts (see Methods for details). Figure~\ref{fig:fractal_dimension}(b) shows the resulting $D(\theta)$ for $\lambda=1, 4,$ and $16$. For $\lambda=1$, $D(\theta)$ remains close to unity over a wide range of tilt angles; the slight deviation at $\theta < 0.05\pi$ is likely a numerical artifact arising from the finite system size, as will be elucidated later. For $\lambda=4$, $D(\theta)$ starts at unity at $\theta=0$ and gradually decreases to a minimum of approximately 0.8 near $\theta=\pi/4$. Notably, this profile exhibits repeated sharp peaks at commensurate angles where periodicity is partially restored. For $\lambda=16$, $D(\theta)$ follows a non-monotonic trend with a minimum of 0.9 at $\theta = 0.08\pi$. Compared to the $\lambda=4$ case, the local enhancements at commensurate angles are less pronounced, rendering the curve significantly smoother. The near-unity $D(\theta)$ at $\lambda=1$ confirms the absence of gap formation between subbands, whereas the values strictly less than unity for $\lambda=4$ and $16$ (ranging from \(0.8\) to \(1.0\) and \(0.9\) to \(1.0\), respectively) signify clear spectral fragmentation. These values remain substantially larger than those of the canonical Harper model [Fig.~\ref{fig:fractal_dimension}(a)], indicating that the fragmentation in the TQM is relatively weaker. These findings are consistent with our earlier observations of the continuous spectrum for $\lambda=1$ and self-similar subband structures for $\lambda=4$ and $16$ [Fig.~\ref{fig:all_HB}].

To verify the fractal dimension calculated from the TQM, we repeat the box-counting analysis using the isospectral dual representation [Eq.~\eqref{eq:isodual}] of the effective Harper-like model.
This dual model preserves the spectrum exactly while reducing the problem to one dimension, it allows simulations at substantially larger system sizes than the original TQM, thereby mitigating finite-size effects.
For $\lambda=1$, $D(\theta)$ remains nearly unity across the entire range, with only a shallow dip near $\theta \approx 0.01\pi$.
This behavior stands in clear contrast to the substantial deviation from unity observed in the TQM at small angles [Fig.~\ref{fig:fractal_dimension}(b)], suggesting that the latter is likely a finite-size system artifact, as mentioned before.
For $\lambda=4$, the $D(\theta)$ values from both models closely align for $\theta \geq 0.08\pi$, whereas the Harper-like model yields larger values than the original TQM in the $\theta < 0.08\pi$ regime. In the $\lambda=16$ case, even better agreement is observed throughout the entire range, with only a minor divergence appearing for $\theta < 0.04\pi$. We attribute these discrepancies to sparse spectral lines and artificial gaps that emerge due to finite-size constraints. Specifically, such features are prominently observed in the low-angle regime within the energy ranges of $4 \lesssim |E| \lesssim 5$ for $\lambda=1$ ($\theta < 0.05\pi$) [Fig.~\ref{fig:all_HB}(d)], $5 \lesssim |E| \lesssim 8$ for $\lambda=4$ ($\theta < 0.1\pi$) [Fig.~\ref{fig:all_HB}(e)], and $|E| \approx 18$ for $\lambda=16$ ($\theta < 0.03\pi$) [Fig.~\ref{fig:all_HB}(f)]. Importantly, these artificial features disappear in the effective model, confirming that their presence in the TQM stems from finite-size effects (see Supplementary Note 3 for a detailed comparison). Since the effective model provides a more faithful representation due to the significantly larger accessible system sizes, its results are considered more robust in the small-$\theta$ limit. Consequently, we conclude that the unity fractal dimension for $\lambda=1$ and the less-than-unity values for $\lambda=4$ and $16$ are generic properties of the system, while their detailed evolution as a function of tilt angle is more accurately captured by the refined results in Fig.~\ref{fig:fractal_dimension}(d). These distinctive angle-dependent patterns of $D(\theta)$ serve as quantitative experimental signatures of HBs arising from the TQM.

We lastly investigate the evolution of the fractal dimension of the TQM across a wide range of the onsite potential amplitude $\lambda$.
Figure~\ref{fig:fractal_dimension}(c) displays the angle-averaged fractal dimension $D_0 = \langle D(\theta) \rangle$ as a function of $\lambda$.
Initially, $D_0$ decreases as $\lambda$ increases, exhibiting a noticeable deviation from unity at $\lambda=2$ where \(D_0 = 0.988\).
This reduction suggests that a fragmented subband structure has become predominant, and our visual inspection into the corresponding energy spectrum supports this prediction (see Supplementary Note 4). 
Consequently, we identify $\lambda=2$ as the threshold for the onset of spectral fragmentation and the emergence of HBs.
The $D_0$ value reaches its minimum of \(D_{0} = 0.816\) at \(\lambda=6\).
We mention that at this point, the subband width is maximized, facilitating the experimental measurement of the subband structure in the HB.
After this minimum, the trend transitions into an upturn, reaching \(D_{0} = 0.927\) at the largest considered \(\lambda=20\).
We speculate that this non-monotonic behavior stems from the different scaling of the subband gap size and the total bandwidth.
While the total bandwidth increases linearly with $\lambda$, the subband gap size grows sharply at first and saturates thereafter; this disproportionate scaling effectively reduces the fractal dimension of the spectrum.
We emphasize that the minimum value, $D_0=0.816$, remains significantly higher than the Harper model's value of \(D_{0} = 0.46\), indicating that a characteristic weak spectral fragmentation persists across the entire range of $\lambda$ values.


\section{Proposal for experimental setup} \label{sec:experimental_setup}

\begin{figure}[t]
    \centering
    \includegraphics[width=\linewidth]{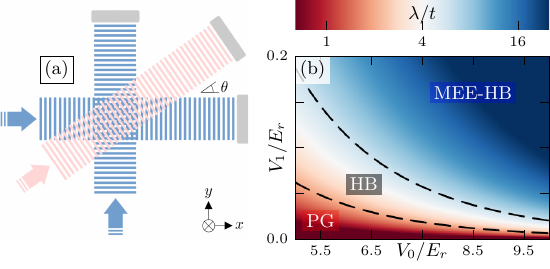}
    \caption{
        (a) Proposed experimental setup for the optical lattice. 
        Two orthogonal standing-wave laser beams interfere to form a primary square lattice, while a third beam, incident at a tilt angle $\theta$, introduces a quasiperiodic onsite potential. 
        (b) Phase diagram in the ($V_0/E_r, V_1/E_r$) plane. 
        The color scale represents the logarithmic ratio $\ln(\lambda/t)$.
        The dashed lines indicate the crossover at \(\lambda/t=2\) and \(\lambda/t=6\).
        The red region indicates the PG regime with \(\lambda/t < 2\), the intermediate region indicates traditional HB regime with \(2 \leq \lambda/t <6\), and the blue region corresponds to the MEE-HB phase in the strong-potential regime (\(\lambda/t \gg 6\)).
    }
    \label{fig:experiment_setup}
\end{figure} 


The model presented here is well-suited for experimental realization using cold atoms in optical lattices \cite{lohse2018exploring, PhysRevLett.111.185301, PhysRevLett.122.110404}.
Specifically, we propose an optical lattice configuration generated by three independent standing waves, all sharing a common wave vector \(k_{L} = 2\pi/a\), where \(a\) denotes the target lattice constant.
Two of these standing waves are aligned orthogonally along the \(x\) and \(y\) axes to create the primary square lattice, while the third is oriented at a tilt angle \(\theta\) to introduce the onsite potential [Fig.~\ref{fig:experiment_setup}(a)].
The non-interacting single-particle Hamiltonian in the continuum for this system is given by:
\begin{equation}
    \left(-\frac{\hbar^2\nabla^2}{2m} + V_\text{lat}(\bm{r}) + V_\text{mod}(\bm{r})\right) \Psi(\bm{r}) = E \Psi(\bm{r}),
\end{equation}
where \(\bm{r}=(x,y)\) represents the continuum spatial coordinates.
The optical potential \(V_\text{lat}(\bm{r}) = V_0[\cos(k_L x) + \cos(k_L y)]\) forms the primary square lattice, while the onsite potential is defined as \(V_\text{mod}(\bm{r}) = V_1 \cos(k_L \bm{n}\cdot \bm{r})\), with the unit vector \(\bm{n}=(\cos\theta,\sin\theta)\).

When the potential depth of the primary lattice \(V_0 \) is sufficiently large compared to both the recoil energy \(E_{r} = \hbar^{2} k_{L}^{2}/(2m)\) and the depth of the onsite potential \(V_1\), the system is accurately described by a tight-binding model ~\cite{PhysRevA.80.021603} as presented in Eq.~\eqref{eq:H}.
In the deep-lattice regime \(V_{0}/E_{r} \gtrsim 5\) considered here, the effective hopping \(t\) and the potential strength \(\lambda\) are given by ~\cite{PhysRevA.80.021603}:
\begin{equation}
\begin{aligned}
    t &\approx \frac{4}{\sqrt{\pi}}E_r\left(\frac{2V_0}{E_r}\right)^{\frac{3}{4}} \exp\left(-2\sqrt{\frac{2V_0}{E_r}}\right), \\
    \lambda &\approx V_1 \exp\left(-\sqrt{\frac{E_r}{2V_0}}\right).
\end{aligned}
\end{equation}
The system exhibits three distinct regimes depending on the ratio \(\lambda/t\).
For \(\lambda/t < 2\), the system remains in the PG phase.
For \(2 \lesssim \lambda/t < 6\), it enters the conventional HB phase.
For \(\lambda/t \gtrsim 6\), the system further crosses over into the MEE-HB phase.
Figure~\ref{fig:experiment_setup}(b) illustrates the phase diagram of these three regimes as a function of the dimensionless lattice parameters $V_0/E_r$ and $V_1/E_r$.
For a given value of $V_0/E_r$, increasing \(V_1/E_r\) changes the system successively from the PG regime to the HB regime and then to the MEE-HB regime.
Notably, the required threshold value of $V_1/E_r$ decreases continuously as $V_0/E_r$ increases.
This behavior can be understood by considering the competing dependencies of the model parameters: in the deep lattice regime, the hopping amplitude \(t\) decreases exponentially with increasing \(V_0/E_r\), whereas the onsite potential strength \(\lambda\) depends only weakly on \(V_0/E_r\).
Consequently, \(\lambda/ t \) scales approximately as
\begin{gather}
    \frac{\lambda}{t} \sim \frac{V_{1}}{E_{r}} \exp\left(2\sqrt{\frac{2V_0}{E_r}}\right).
\end{gather}
Solving the conditions \(\lambda/t \approx 2\) and \(\lambda/t \approx 6\) yields the corresponding threshold values of \(V_{1}^{*}\) for the PG-to-HB and HB-to-MEE-HB crossovers, respectively.
Both thresholds decrease rapidly as \(V_0/E_r\) increases, with the same leading exponential dependence:
\begin{gather}
    V_{1}^{\ast} \sim E_{r} \exp\left(-2\sqrt{ \frac{2V_0}{E_r}}\right), 
\end{gather}
up to different numerical prefactors associated with the two phase boundaries.
This analytical scaling closely matches the numerical crossover lines indicated by the black dashed lines in Fig.~\ref{fig:experiment_setup}(b).
This makes the deep-lattice regime particularly favorable for experimental access to both the HB and MEE-HB without requiring a large onsite potential amplitude $V_1$.

\section{Alternative routes to MEE-HB} \label{sec:other_possible_routes}

\begin{figure}
    \centering
    \includegraphics[width=\columnwidth]{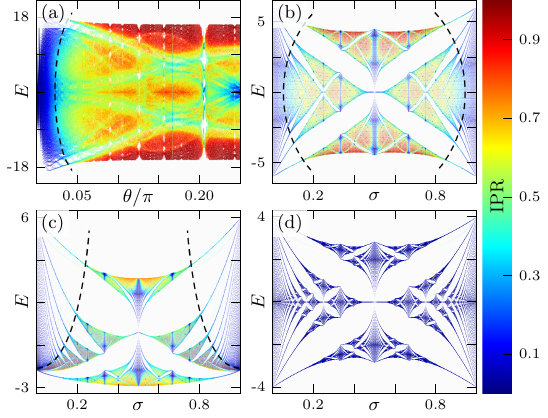}
    \caption{
    \textbf{HBs from other models}.
    (a) Energy spectrum of the bidirectional potential model as a function of the tilt angle \(\theta\) at \(\lambda = 8\) with 10 phase realizations. 
    Energy spectra from (b) Anisotropic Harper model, (c) extended Harper model featuring next-nearest-neighbor hopping, and (d) standard isotropic Harper model, each plotted as a function of the potential wave vector \(\alpha\).
    Taking states with \({\rm IPR} > 0.3\) as localized, panel~(a) displays MEs without a signature of HB hierarchy, whereas panels~(b) and (c) exhibit MEE-HBs.
    In contrast, (d) shows the conventional HB, which lacks MEs.
    The result in (a) is obtained for a \(30\times30\) square lattice under OBCs, while those in (b)--(d) are obtained for the one-dimensional lattices of size \(L=500\) under OBCs by scanning 1,111 uniformly spaced values of \(\sigma\in[0,1]\).
    }
    \label{fig:other_butterfly}
\end{figure}

To the best of our knowledge, the emergence of MEs within an HB has not been well documented in the literature.
This motivates us to explore alternative configurations for realizing the MEE-HB and to compare our findings with these scenarios.
Furthermore, this comparative analysis clarifies the key ingredients required to generate an MEE-HB, particularly by contrasting the TQM with these alternatives.
As a first comparison, we consider a bidirectional potential extension of our TQM.
In this case, the kinetic terms remain identical to Eq.~\eqref{eq:schrodinger_eqn}, but the onsite potential is replaced by
\begin{align}
    V_{n,m} = & \; \lambda \cos[2\pi (n\cos{\theta}+m\sin{\theta})] \notag\\[3pt]
    & + \lambda \cos[2\pi (-n\sin{\theta}+m\cos{\theta})].
\end{align}
Figure~\ref{fig:other_butterfly}(a) displays the resulting energy spectra as a function of the tilt angle \(\theta\) obtained by numerically solving the corresponding tight-binding model.
Although this bidirectional setup supports MEs, it lacks the characteristic HB hierarchy and therefore fails to realize the MEE-HB structure found in our original model.
The origin of this difference becomes clear in the tilted coordinate basis \((\ell,j)\), where the Schr\"odinger equation takes the following form,
\begin{align}
    E\psi_{\ell,j} = & \; \lambda\cos(2\pi\alpha\ell)\psi_{\ell,j} + \lambda\cos(2\pi\alpha j)\psi_{\ell,j} \\[3pt]
    & + \psi_{\ell+q,j-p} + \psi_{\ell-q,j+p} + \psi_{\ell+p,j+q} + \psi_{\ell-p,j-q}. \notag
\end{align}
In this basis, the potential depends on both coordinates $\ell$ and $j$, giving rise to quasiperiodicity along two orthogonal directions. The quasiperiodic potential along the $j$-direction prevents the reduction of the system to the decoupled 1D equations derived in Eq.~\eqref{eq:effective_1d}. Instead, applying the Fourier transform as in Eq.~\eqref{eq:bloch_j} introduces a coupling between equations with different wave vectors.
This indicates that the self-recursive structure inherent to the 1D Harper equation---which is fundamental to fractal energy splitting---is broken by the additional quasiperiodic dimensionality. This suggests that the 1D orientation of the quasiperiodic potential in our original model is the crucial ingredient for allowing the fractal butterfly structure to coexist with an ME.

The second scenario we consider is a system of non-interacting particles on a square lattice subjected to a uniform perpendicular magnetic field~\cite{hofstadter1976energy}.
The corresponding Schrödinger equation is given by
\begin{align}
    E\psi_{n,m} = & \; t_x(\psi_{n+1,m} + \psi_{n-1,m}) \notag \\[3pt]
                  & + t_y (e^{-i 2\pi\sigma n}\psi_{n,m+1} + e^{i 2\pi\sigma n}\psi_{n,m-1}), \label{eq:AA_model}
\end{align}
where \(t_{x}\) and \(t_{y}\) denote the nearest-neighbor hopping amplitudes along the \(x\)- and \(y\)-directions, respectively, and \(\sigma = \phi/\phi_{0}\) represents the magnetic flux per unit cell in units of the flux quantum \(\phi_{0}\).
By employing the plane-wave ansatz \(\psi_{n,m} = g_{n} \exp(i\nu m)\), where \(\nu\) is the wave vector along the \(y\)-direction, and setting \(t_{x} = 1\) with \(\lambda = t_y/t_x\), we obtain an effective 1D Harper equation in Eq.~\eqref{eq:harper}.
Figure~\ref{fig:other_butterfly}(b) displays the energy spectra as a function of \(\sigma\), obtained by numerically solving Eq.~\eqref{eq:harper} for \(\lambda=2\).
Notably, the HB of this model displays both extended states (\(\text{IPR} \leq 0.3 \); blue) and high-IPR line-like localized states (\(\text{IPR} > 0.3\); red) in the two-dimensional system in Eq.~\eqref{eq:AA_model}, separated by apparent MEs (\(\text{IPR} = 0.3\); black lines).
This behavior stands in contrast to the standard isotropic case (\(\lambda=1\)), where the MEs are absent and all states remain extended [Fig.~\ref{fig:other_butterfly}(d)].
Consequently, the anisotropic model provides an alternative method for generating an apparent MEE-HB in a finite system. 

As a third scenario, we consider non-interacting particles with isotropic nearest-neighbor hopping, \(t_{x} = t_{y} = t\), supplemented by an additional next-nearest-neighbor hopping amplitude \(t^{\prime}\).
The corresponding Schr\"odinger equation is given by
\begin{align}
    E \psi_{n,m} = &\; t(\psi_{n+1,m} + \psi_{n-1,m}) \notag\\[3pt]
                   & + te^{-i2\pi\sigma n}\psi_{n,m+1} + te^{+i2\pi\sigma n}\psi_{n,m-1} \notag\\[3pt]
                   & + t^{\prime} e^{-i2\pi\sigma n}\psi_{n+1,m+1} + t^{\prime} e^{+i2\pi\sigma n}\psi_{n+1,m-1} \notag\\[3pt]
                   & + t^{\prime} e^{-i2\pi\sigma (n-1)}\psi_{n-1,m+1} \notag\\[3pt]
                   & + t^{\prime} e^{+i2\pi\sigma (n-1)}\psi_{n-1,m-1}
\end{align}
Using the same plane-wave ansatz \(\psi_{n,m}= g_{n}\exp(i\nu m)\), and setting \(t=1\) and \(\lambda = t^{\prime}/t\), we obtain the following effective 1D Harper-like equation:
\begin{align} \label{eq:NNN_Harper_eq}
    Eg_n = & \; 2\cos(2\pi \sigma n - \nu)g_n \notag\\[3pt]
    & + \left[1 + 2\lambda\cos(2\pi \sigma n -\nu)\right] g_{n+1} \notag\\[3pt]
    & + \left[1 + 2\lambda\cos(2\pi \sigma (n-1) -\nu)\right] g_{n-1}.
\end{align}
This model is a variant of the extended-Harper class~\cite{Avila2017} with modified hopping coefficients.
Figure~\ref{fig:other_butterfly}(c) illustrates the energy spectra as a function of \(\sigma\) for \(\lambda=0.5\), obtained by numerically solving Eq.~\eqref{eq:NNN_Harper_eq}.
In this extended model, the HB exhibits a coexistence of extended states (\(\text{IPR} \leq 0.3 \); blue) and high-IPR line-like localized states (\(\text{IPR} > 0.3\); red), which are separated by distinct apparent MEs (\(\text{IPR} =  0.3\); black lines).
Our numerical results indicate that the regime of localized states reaches its maximum extent at \(\lambda=0.5\) and systematically contracts as \(\lambda\) deviates from this value.
These observations demonstrate that tuning the ratio between nearest-neighbor and next-nearest-neighbor hopping amplitudes provides a robust mechanism for generating an apparent MEE-HB in a finite system.

Our analysis based on the second and third scenarios explicitly demonstrates that an MEE-HB can be realized in a traditional square lattice system under a uniform perpendicular magnetic field in a finite system.
However, in the context of optical lattices, achieving such a realization typically necessitates the implementation of synthetic magnetic fields via complex laser configurations or Raman-assisted tunneling~\cite{Jotzu2014, PhysRevLett.111.185301, PhysRevLett.111.185302}, both of which remain experimentally challenging due to heating and technical complexities~\cite{RevModPhys.83.1523}.
Furthermore, engineering highly anisotropic hopping ratios or next-nearest-neighbor amplitudes---particularly those approaching half the magnitude of nearest-neighbor strengths---requires sophisticated multi-wave vector laser setups, posing significant additional hurdles~\cite{Tarruell2012, Jotzu2014}.
In contrast, the tilt-induced quasiperiodic onsite potential achieves these two essential effects using only three laser beams of the same wave vector.
This provides a significantly more streamlined platform for investigating the interplay between fractal energy splitting and localization transitions, namely an MEE-HB.

\section{Conclusion} \label{sec:discussion}

In summary, we have demonstrated that a tilt-induced 1D quasiperiodic potential in a 2D square lattice provides a unified framework for investigating the interplay between fractal energy splitting and mobility edge phenomena. Through numerical diagonalization of the tight-binding Hamiltonian, we mapped the resulting Hofstadter butterfly as a function of the tilt angle and energy. While the fractal spectrum consists of fully delocalized states in the weak-potential regime, it develops energy-dependent mobility edges as the potential strength increases—a configuration that remains notably elusive in standard periodic and simple quasiperiodic models. Our derivation of an effective 1D Harper-like equation reveals that the emergence of this Hofstadter butterfly is governed by a self-recursive fractal relationship analogous to the standard Harper equation. Crucially, the coexistence of multi-range hopping terms inherent in our 2D-to-1D mapping provides the underlying mechanism for the mobility edge features. Furthermore, we proposed specific optical lattice experimental setups to observe this mobility-edge-embedded Hofstadter butterfly. This approach offers significant practical advantages, as it bypasses the need for complex experimental prerequisites such as fine-tuned next-nearest-neighbor hopping or extreme hopping anisotropy. Finally, the high fractal dimension of this spectrum provides a robust spectral density, serving as a distinct experimental signature that differentiates it from the sparser butterflies typically found in magnetic-field-based systems.

\section{Methods}

\subsection{Lorentzian-broaden density of states}

To obtain the DOS maps in the \((\theta, E)\) plane shown in Figs.~\ref{fig:all_HB}(a)--(c), we broadened the discrete eigenvalues at each tilt angle \(\theta\) with a Lorentzian kernel.
For each \(\theta\), we diagonalized the TQM tight-binding Hamiltonian under OBCs to obtain the full set of eigenvalues \(\{E_{\mu}(\theta)\}\).
For each \(\theta\), the DOS was evaluated on a uniform energy grid covering the full spectrum.
Each discrete eigenvalue was then broadened by a Lorentzian kernel according to
\begin{gather}
    \rho(E;\theta) = \frac{1}{N_{\mathrm{tot}}} \sum_{\mu=1}^{N_{\mathrm{tot}}} \frac{1}{\pi} \frac{\eta}{\left(E-E_\mu(\theta)\right)^2+\eta^2},
\end{gather}
where \(N_{\mathrm{tot}} = L^{2}\) is the total number of states and \(\eta\) is the Lorentzian broadening width.
The factor \(1/N_{\mathrm{tot}}\) normalizes the DOS such that the following condition holds for each \(\theta\):
\begin{gather}
    \int_{\mathbb{R}} dE\, \rho(E;\theta)=1
\end{gather}
In the numerical implementation, we used 1200 samples of \(\theta\) and a broadening width of \(\eta = 1.2\).

\subsection{Numerical procedures for spectral calculations}

To obtain the DOS and IPR maps shown in Figs.~\ref{fig:all_HB}(d)--(f), and to construct the spectral datasets used for the fractal-dimension analysis in Figs.~\ref{fig:fractal_dimension}(b) and (c), we diagonalized the TQM Hamiltonian in Eq.~\eqref{eq:schrodinger_eqn} on a finite square lattice.
The calculation was performed on an \(80\times 80\) square lattice with OBCs.
The tilt angle was sampled over 1200 uniformly spaced values in the interval \([0,\pi/4]\).
For each value of \(\theta\), we obtained the full spectrum and the corresponding IPR values by exact diagonalization.

To obtain the spectra shown in Figs.~\ref{fig:all_HB}(g)--(i), we diagonalized the finite Hamiltonian associated with the effective Harper-like model in Eq.~\eqref{eq:effective_1d}.
The calculation was performed for rational slopes \(\tan\theta = p/q\), where \(p\) and \(q\) are coprime integers satisfying \(p^{2} + q^{2} \leq 121^{2}\).
In constructing the rational-slope dataset, we excluded Pythagorean pairs satisfying \(p^{2} + q^{2} = r^{2}\) with integer \(r\), so that the modulation frequency \(\alpha\) remained incommensurate.
The tilt angle corresponding to each \((p,q)\) pair was identified as \(\theta = \arctan(p/q)\).
For each \((p,q)\), we diagonalized the finite Hamiltonian associated with Eq.~\eqref{eq:effective_1d} and obtained its full set of eigenvalues.
In principle, the finite-size spectrum can depend on the transverse Bloch momentum \(\nu\).
However, we confirmed that, for sufficiently large system size, the overall spectral structure became insensitive to \(\nu\), except at \(\theta = \pi/4\).
In the numerical implementation, we used \(\nu = 0\) and system size \(800\).
For the fractal-dimension analysis in Fig.~\ref{fig:fractal_dimension}(d), we used the dual representation of the effective Harper-like model with a larger system size \(6400\) to improve spectral resolution.
A more detailed analysis of the \(\nu\)-dependence is provided in Supplementary Note 2.

\subsection{Box counting method}

To obtain the fractal dimension shown in Fig.~\ref{fig:fractal_dimension}, we applied a box-counting analysis to both the Harper spectrum and the TQM spectra.
For the TQM, the spectra were computed on \(80\times80\) square lattices under open boundary conditions over 1200 uniformly sampled angles in the range \(\theta\in[0,\pi/4]\).
Before box counting, the TQM spectra were filtered to remove states with large weights near the sample edge, using a boundary layer two lattice sites thick.
For each retained spectrum, the energy levels were rescaled to the unit interval \([0,1]\), and near-duplicate levels were removed using a numerical tolerance of \(10^{-10}\). 
We then covered the normalized energy axis with boxes of size \(\epsilon\), where \(\epsilon\) denotes the box size in the rescaled spectrum, and counted the number \(N(\epsilon)\) of boxes containing at least one energy level.
The box count was evaluated over a logarithmically spaced set of box sizes, using origin averaging over eight shifted grids to reduce grid dependence.
The fitting range was chosen by $4<N(\epsilon)<0.9N_0$, where $N_0$ represents the number of distinct energy levels (after elimination of near-duplicate levels).
For both the Harper and TQM spectra, we used \(\epsilon_{\rm base}=2.0\) and 20 logarithmically spaced box sizes over the range \(\epsilon\in[2^{-11},2^{-2}]\). 
The fractal dimension \(D\) was extracted by a linear fit to \(\ln N(\epsilon)\) versus \(\ln(1/\epsilon)\).
We adopted \(\epsilon_{\rm base}=2.0\) in the main text, since it yielded the most stable scaling estimates for the TQM.
Results for other values of \(\epsilon_{\rm base}\) and additional numerical details are provided in Supplementary Note 3.

\begin{acknowledgments}
This research was supported by an appointment to the JRG Program at the APCTP through the Science and Technology Promotion Fund and Lottery Fund of the Korean Government, the Korean Local Governments (Gyeongsangbuk-do Province and Pohang City), and the National Research Foundation of Korea (NRF) funded by the Korean government (Ministry of Science and ICT, MSIT) (No. RS-2026-25499525).
\end{acknowledgments}

\bibliography{ref}

@Article{Dean2013,
author={Dean, C. R.
and Wang, L.
and Maher, P.
and Forsythe, C.
and Ghahari, F.
and Gao, Y.
and Katoch, J.
and Ishigami, M.
and Moon, P.
and Koshino, M.
and Taniguchi, T.
and Watanabe, K.
and Shepard, K. L.
and Hone, J.
and Kim, P.},
title={Hofstadter's butterfly and the fractal quantum Hall effect in moir{\'e} superlattices},
journal={Nature},
year={2013},
month={May},
day={01},
volume={497},
number={7451},
pages={598-602},
issn={1476-4687},
doi={10.1038/nature12186},
url={https://doi.org/10.1038/nature12186}
}

@article{kraus2013four,
title = {Four-Dimensional Quantum Hall Effect in a Two-Dimensional Quasicrystal},
author = {Kraus, Yaacov E. and Ringel, Zohar and Zilberberg, Oded},
journal = {Phys. Rev. Lett.},
volume = {111},
issue = {22},
pages = {226401},
numpages = {5},
year = {2013},
month = {Nov},
publisher = {American Physical Society},
doi = {10.1103/PhysRevLett.111.226401},
url = {https://link.aps.org/doi/10.1103/PhysRevLett.111.226401}
}

@Article{lohse2018exploring,
author={Lohse, Michael
and Schweizer, Christian
and Price, Hannah M.
and Zilberberg, Oded
and Bloch, Immanuel},
title={Exploring 4D quantum Hall physics with a 2D topological charge pump},
journal={Nature},
year={2018},
month={Jan},
day={01},
volume={553},
number={7686},
pages={55-58},
issn={1476-4687},
doi={10.1038/nature25000},
url={https://doi.org/10.1038/nature25000}
}

@article{luschen2018single,
  title = {Single-Particle Mobility Edge in a One-Dimensional Quasiperiodic Optical Lattice},
  author = {L\"uschen, Henrik P. and Scherg, Sebastian and Kohlert, Thomas and Schreiber, Michael and Bordia, Pranjal and Li, Xiao and Das Sarma, S. and Bloch, Immanuel},
  journal = {Phys. Rev. Lett.},
  volume = {120},
  issue = {16},
  pages = {160404},
  numpages = {6},
  year = {2018},
  month = {Apr},
  publisher = {American Physical Society},
  doi = {10.1103/PhysRevLett.120.160404},
  url = {https://link.aps.org/doi/10.1103/PhysRevLett.120.160404}
}

@article{An2021InteractionsMobilityEdges,
  author  = {An, Fangzhao Alex and Padavić, Kevin and Scherg, Stefan and Janot, Alexandre and Gadway, Bryce},
  title   = {Interactions and Mobility Edges: Observing the Generalized Aubry–André Model},
  journal = {Phys. Rev. Lett.},
  year    = {2021},
  volume  = {126},
  number  = {4},
  pages   = {040603},
  doi     = {10.1103/PhysRevLett.126.040603}
}

@article{SOKOLOFF1985189,
title = {Unusual band structure, wave functions and electrical conductance in crystals with incommensurate periodic potentials},
journal = {Phys. Rep.},
volume = {126},
number = {4},
pages = {189-244},
year = {1985},
issn = {0370-1573},
doi = {https://doi.org/10.1016/0370-1573(85)90088-2},
url = {https://www.sciencedirect.com/science/article/pii/0370157385900882},
author = {J.B Sokoloff}
}

@article{Macia_2006,
doi = {10.1088/0034-4885/69/2/R03},
url = {https://doi.org/10.1088/0034-4885/69/2/R03},
year = {2005},
month = {dec},
publisher = {},
volume = {69},
number = {2},
pages = {397},
author = {Maciá, Enrique},
title = {The role of aperiodic order in science and technology},
journal = {Rep. Prog. Phys.}
}

@article{PhysRevLett.109.106402,
  title = {Topological States and Adiabatic Pumping in Quasicrystals},
  author = {Kraus, Yaacov E. and Lahini, Yoav and Ringel, Zohar and Verbin, Mor and Zilberberg, Oded},
  journal = {Phys. Rev. Lett.},
  volume = {109},
  issue = {10},
  pages = {106402},
  numpages = {5},
  year = {2012},
  month = {Sep},
  publisher = {American Physical Society},
  doi = {10.1103/PhysRevLett.109.106402},
  url = {https://link.aps.org/doi/10.1103/PhysRevLett.109.106402}
}

@article{PhysRevLett.111.185302,
  title = {Realizing the Harper Hamiltonian with Laser-Assisted Tunneling in Optical Lattices},
  author = {Miyake, Hirokazu and Siviloglou, Georgios A. and Kennedy, Colin J. and Burton, William Cody and Ketterle, Wolfgang},
  journal = {Phys. Rev. Lett.},
  volume = {111},
  issue = {18},
  pages = {185302},
  numpages = {5},
  year = {2013},
  month = {Oct},
  publisher = {American Physical Society},
  doi = {10.1103/PhysRevLett.111.185302},
  url = {https://link.aps.org/doi/10.1103/PhysRevLett.111.185302}
}

@article{PhysRevLett.111.185301,
  title = {Realization of the Hofstadter Hamiltonian with Ultracold Atoms in Optical Lattices},
  author = {Aidelsburger, M. and Atala, M. and Lohse, M. and Barreiro, J. T. and Paredes, B. and Bloch, I.},
  journal = {Phys. Rev. Lett.},
  volume = {111},
  issue = {18},
  pages = {185301},
  numpages = {5},
  year = {2013},
  month = {Oct},
  publisher = {American Physical Society},
  doi = {10.1103/PhysRevLett.111.185301},
  url = {https://link.aps.org/doi/10.1103/PhysRevLett.111.185301}
}

@article{PhysRevB.14.2239,
  title = {Energy levels and wave functions of Bloch electrons in rational and irrational magnetic fields},
  author = {Hofstadter, Douglas R.},
  journal = {Phys. Rev. B},
  volume = {14},
  issue = {6},
  pages = {2239--2249},
  numpages = {0},
  year = {1976},
  month = {Sep},
  publisher = {American Physical Society},
  doi = {10.1103/PhysRevB.14.2239},
  url = {https://link.aps.org/doi/10.1103/PhysRevB.14.2239}
}

@article{PhysRevLett.120.160404,
  title = {Single-Particle Mobility Edge in a One-Dimensional Quasiperiodic Optical Lattice},
  author = {L\"uschen, Henrik P. and Scherg, Sebastian and Kohlert, Thomas and Schreiber, Michael and Bordia, Pranjal and Li, Xiao and Das Sarma, S. and Bloch, Immanuel},
  journal = {Phys. Rev. Lett.},
  volume = {120},
  issue = {16},
  pages = {160404},
  numpages = {6},
  year = {2018},
  month = {Apr},
  publisher = {American Physical Society},
  doi = {10.1103/PhysRevLett.120.160404},
  url = {https://link.aps.org/doi/10.1103/PhysRevLett.120.160404}
}

@article{PhysRevLett.61.2144,
  title = {Mobility Edge in a Model One-Dimensional Potential},
  author = {Das Sarma, S. and He, Song and Xie, X. C.},
  journal = {Phys. Rev. Lett.},
  volume = {61},
  issue = {18},
  pages = {2144--2147},
  numpages = {0},
  year = {1988},
  month = {Oct},
  publisher = {American Physical Society},
  doi = {10.1103/PhysRevLett.61.2144},
  url = {https://link.aps.org/doi/10.1103/PhysRevLett.61.2144}
}

@article{PhysRevLett.122.110404,
  title = {Matter-Wave Diffraction from a Quasicrystalline Optical Lattice},
  author = {Viebahn, Konrad and Sbroscia, Matteo and Carter, Edward and Yu, Jr-Chiun and Schneider, Ulrich},
  journal = {Phys. Rev. Lett.},
  volume = {122},
  issue = {11},
  pages = {110404},
  numpages = {6},
  year = {2019},
  month = {Mar},
  publisher = {American Physical Society},
  doi = {10.1103/PhysRevLett.122.110404},
  url = {https://link.aps.org/doi/10.1103/PhysRevLett.122.110404}
}

@article{PhysRevA.80.021603,
  title = {Localization in one-dimensional incommensurate lattices beyond the Aubry-Andr\'e model},
  author = {Biddle, J. and Wang, B. and Priour, D. J. and Das Sarma, S.},
  journal = {Phys. Rev. A},
  volume = {80},
  issue = {2},
  pages = {021603},
  numpages = {4},
  year = {2009},
  month = {Aug},
  publisher = {American Physical Society},
  doi = {10.1103/PhysRevA.80.021603},
  url = {https://link.aps.org/doi/10.1103/PhysRevA.80.021603}
}

@article{RevModPhys.83.1523,
  title = {Colloquium: Artificial gauge potentials for neutral atoms},
  author = {Dalibard, Jean and Gerbier, Fabrice and Juzeli\ifmmode \bar{u}\else \={u}\fi{}nas, Gediminas and \"Ohberg, Patrik},
  journal = {Rev. Mod. Phys.},
  volume = {83},
  issue = {4},
  pages = {1523--1543},
  numpages = {0},
  year = {2011},
  month = {Nov},
  publisher = {American Physical Society},
  doi = {10.1103/RevModPhys.83.1523},
  url = {https://link.aps.org/doi/10.1103/RevModPhys.83.1523}
}

@Article{Tarruell2012,
author={Tarruell, Leticia
and Greif, Daniel
and Uehlinger, Thomas
and Jotzu, Gregor
and Esslinger, Tilman},
title={Creating, moving and merging Dirac points with a Fermi gas in a tunable honeycomb lattice},
journal={Nature},
year={2012},
month={Mar},
day={01},
volume={483},
number={7389},
pages={302-305},
issn={1476-4687},
doi={10.1038/nature10871},
url={https://doi.org/10.1038/nature10871}
}

@Article{Jotzu2014,
author={Jotzu, Gregor
and Messer, Michael
and Desbuquois, R{\'e}mi
and Lebrat, Martin
and Uehlinger, Thomas
and Greif, Daniel
and Esslinger, Tilman},
title={Experimental realization of the topological Haldane model with ultracold fermions},
journal={Nature},
year={2014},
month={Nov},
day={01},
volume={515},
number={7526},
pages={237-240},
issn={1476-4687},
doi={10.1038/nature13915},
url={https://doi.org/10.1038/nature13915}
}

@article{lee2026structural,
title = {Structural constraints on mobility edges in one-dimensional quasiperiodic systems},
author = {Lee, Sanghoon and \ifmmode \check{C}\else \v{C}\fi{}ade\ifmmode \check{z}\else \v{z}\fi{}, Tilen and Kim, Kyoung-Min},
journal = {Phys. Rev. B},
volume = {113},
issue = {17},
pages = {174202},
numpages = {8},
year = {2026},
month = {May},
publisher = {American Physical Society},
doi = {10.1103/pyj9-ns4f},
url = {https://link.aps.org/doi/10.1103/pyj9-ns4f}
}

@article{hofstadter1976energy,
  title = {Energy levels and wave functions of Bloch electrons in rational and irrational magnetic fields},
  author = {Hofstadter, Douglas R.},
  journal = {Phys. Rev. B},
  volume = {14},
  issue = {6},
  pages = {2239--2249},
  numpages = {0},
  year = {1976},
  month = {Sep},
  publisher = {American Physical Society},
  doi = {10.1103/PhysRevB.14.2239},
  url = {https://link.aps.org/doi/10.1103/PhysRevB.14.2239}
}

@Article{Avila2017,
author={Avila, A.
and Jitomirskaya, S.
and Marx, C. A.},
title={Spectral theory of extended Harper's model and a question by Erd{\H{o}}s and Szekeres},
journal={Invent. math.},
year={2017},
month={Oct},
day={01},
volume={210},
number={1},
pages={283-339},
issn={1432-1297},
doi={10.1007/s00222-017-0729-1},
url={https://doi.org/10.1007/s00222-017-0729-1}
}

@article{LIEBOVITCH1989386,
title = {A fast algorithm to determine fractal dimensions by box counting},
journal = {Phys. Lett. A},
volume = {141},
number = {8},
pages = {386-390},
year = {1989},
issn = {0375-9601},
doi = {https://doi.org/10.1016/0375-9601(89)90854-2},
url = {https://www.sciencedirect.com/science/article/pii/0375960189908542},
author = {Larry S. Liebovitch and Tibor Toth}
}

@Article{Nuckolls2025,
author={Nuckolls, Kevin P.
and Scheer, Michael G.
and Wong, Dillon
and Oh, Myungchul
and Lee, Ryan L.
and Herzog-Arbeitman, Jonah
and Watanabe, Kenji
and Taniguchi, Takashi
and Lian, Biao
and Yazdani, Ali},
title={Spectroscopy of the fractal Hofstadter energy spectrum},
journal={Nature},
year={2025},
month={Mar},
day={01},
volume={639},
number={8053},
pages={60-66},
issn={1476-4687},
doi={10.1038/s41586-024-08550-2},
url={https://doi.org/10.1038/s41586-024-08550-2}
}

@article{du2018floquet,
title = {Floquet Hofstadter butterfly on the kagome and triangular lattices},
author = {Du, Liang and Chen, Qi and Barr, Aaron D. and Barr, Ariel R. and Fiete, Gregory A.},
journal = {Phys. Rev. B},
volume = {98},
issue = {24},
pages = {245145},
numpages = {12},
year = {2018},
month = {Dec},
publisher = {American Physical Society},
doi = {10.1103/PhysRevB.98.245145},
url = {https://link.aps.org/doi/10.1103/PhysRevB.98.245145}
}

@article{PhysRevLett.86.1062,
  title = {Hofstadter Butterfly and Integer Quantum Hall Effect in Three Dimensions},
  author = {Koshino, M. and Aoki, H. and Kuroki, K. and Kagoshima, S. and Osada, T.},
  journal = {Phys. Rev. Lett.},
  volume = {86},
  issue = {6},
  pages = {1062--1065},
  numpages = {0},
  year = {2001},
  month = {Feb},
  publisher = {American Physical Society},
  doi = {10.1103/PhysRevLett.86.1062},
  url = {https://link.aps.org/doi/10.1103/PhysRevLett.86.1062}
}

@article{Pradhan_2016,
doi = {10.1088/0953-8984/28/50/505502},
url = {https://doi.org/10.1088/0953-8984/28/50/505502},
year = {2016},
month = {oct},
publisher = {IOP Publishing},
volume = {28},
number = {50},
pages = {505502},
author = {Pradhan, Subhasree},
title = {Hofstadter butterfly in the Falicov–Kimball model on some finite 2D lattices},
journal = {J. Phys. Condens. Matter}
}

@article{sun2024nonhermitian,
title = {Non-Hermitian quantum fractals},
author = {Sun, Junsong and Li, Chang-An and Guo, Qingyang and Zhang, Weixuan and Feng, Shiping and Zhang, Xiangdong and Guo, Huaiming and Trauzettel, Bj\"orn},
journal = {Phys. Rev. B},
volume = {110},
issue = {20},
pages = {L201103},
numpages = {6},
year = {2024},
month = {Nov},
publisher = {American Physical Society},
doi = {10.1103/PhysRevB.110.L201103},
url = {https://link.aps.org/doi/10.1103/PhysRevB.110.L201103}
}

@Article{Ponomarenko2013,
author={Ponomarenko, L. A.
and Gorbachev, R. V.
and Yu, G. L.
and Elias, D. C.
and Jalil, R.
and Patel, A. A.
and Mishchenko, A.
and Mayorov, A. S.
and Woods, C. R.
and Wallbank, J. R.
and Mucha-Kruczynski, M.
and Piot, B. A.
and Potemski, M.
and Grigorieva, I. V.
and Novoselov, K. S.
and Guinea, F.
and Fal'ko, V. I.
and Geim, A. K.},
title={Cloning of Dirac fermions in graphene superlattices},
journal={Nature},
year={2013},
month={May},
day={01},
volume={497},
number={7451},
pages={594-597},
issn={1476-4687},
doi={10.1038/nature12187},
url={https://doi.org/10.1038/nature12187}
}

@article{doi:10.1126/science.1237240,
author = {B. Hunt  and J. D. Sanchez-Yamagishi  and A. F. Young  and M. Yankowitz  and B. J. LeRoy  and K. Watanabe  and T. Taniguchi  and P. Moon  and M. Koshino  and P. Jarillo-Herrero  and R. C. Ashoori },
title = {Massive Dirac Fermions and Hofstadter Butterfly in a van der Waals Heterostructure},
journal = {Science},
volume = {340},
number = {6139},
pages = {1427-1430},
year = {2013},
doi = {10.1126/science.1237240},
URL = {https://www.science.org/doi/abs/10.1126/science.1237240},
}

@incollection{Li24,
author = {Larry Li and Marcin Abram and Abhinav Prem and Stephan Haas},
title = {Hofstadter Butterflies in Topological Insulators},
booktitle = {Recent Topics on Topology - From Classical to Modern Applications},
publisher = {IntechOpen},
address = {London},
year = {2024},
editor = {Paul Bracken},
chapter = {1},
doi = {10.5772/intechopen.1006115},
url = {https://doi.org/10.5772/intechopen.1006115}
}

@article{7std-nbqw,
  title = {Exact multiple complex mobility edges and quantum state engineering in coupled one-dimensional quasicrystals},
  author = {Wang, Li and Wang, Zhenbo and Liu, Jiaqi and Chen, Shu},
  journal = {Phys. Rev. B},
  volume = {112},
  issue = {10},
  pages = {104207},
  numpages = {12},
  year = {2025},
  month = {Sep},
  publisher = {American Physical Society},
  doi = {10.1103/7std-nbqw},
  url = {https://link.aps.org/doi/10.1103/7std-nbqw}
}

@article{PhysRevLett.114.146601,
  title = {Nearest Neighbor Tight Binding Models with an Exact Mobility Edge in One Dimension},
  author = {Ganeshan, Sriram and Pixley, J. H. and Das Sarma, S.},
  journal = {Phys. Rev. Lett.},
  volume = {114},
  issue = {14},
  pages = {146601},
  numpages = {5},
  year = {2015},
  month = {Apr},
  publisher = {American Physical Society},
  doi = {10.1103/PhysRevLett.114.146601},
  url = {https://link.aps.org/doi/10.1103/PhysRevLett.114.146601}
}

@article{doi:10.1126/science.1209019,
author = {S. S. Kondov  and W. R. McGehee  and J. J. Zirbel  and B. DeMarco },
title = {Three-Dimensional Anderson Localization of Ultracold Matter},
journal = {Science},
volume = {334},
number = {6052},
pages = {66-68},
year = {2011},
doi = {10.1126/science.1209019},
URL = {https://www.science.org/doi/abs/10.1126/science.1209019},
}

@article{PhysRevLett.126.040603,
  title = {Interactions and Mobility Edges: Observing the Generalized Aubry-Andr\'e Model},
  author = {An, Fangzhao Alex and Padavi\ifmmode \acute{c}\else \'{c}\fi{}, Karmela and Meier, Eric J. and Hegde, Suraj and Ganeshan, Sriram and Pixley, J. H. and Vishveshwara, Smitha and Gadway, Bryce},
  journal = {Phys. Rev. Lett.},
  volume = {126},
  issue = {4},
  pages = {040603},
  numpages = {6},
  year = {2021},
  month = {Jan},
  publisher = {American Physical Society},
  doi = {10.1103/PhysRevLett.126.040603},
  url = {https://link.aps.org/doi/10.1103/PhysRevLett.126.040603}
}

@inproceedings{aubry1980analyticity,
  title={Analyticity breaking and Anderson localization in incommensurate lattices},
  author={Aubry, Serge and Andr{\'e}, Gilles},
  booktitle={Ann. Isr. Phys. Soc.},
  volume={3},
  number={133},
  pages={18},
  year={1980}
}

@book{falconer2014fractal,
author    = {Kenneth Falconer},
title     = {Fractal Geometry: Mathematical Foundations and Applications},
edition   = {3},
publisher = {Wiley},
year      = {2014}
}

@book{vicsek1992fractal,
author    = {Tam{\'a}s Vicsek},
title     = {Fractal Growth Phenomena},
publisher = {World Scientific},
year      = {1992}
}

@article{soukoulis1982localization,
title = {Localization in One-Dimensional Lattices in the Presence of Incommensurate Potentials},
author = {Soukoulis, C. M. and Economou, E. N.},
journal = {Phys. Rev. Lett.},
volume = {48},
issue = {15},
pages = {1043--1046},
numpages = {0},
year = {1982},
month = {Apr},
publisher = {American Physical Society},
doi = {10.1103/PhysRevLett.48.1043},
url = {https://link.aps.org/doi/10.1103/PhysRevLett.48.1043},
}

@article{biddle2009localizationA,
title = {Localization in one-dimensional incommensurate lattices beyond the Aubry-Andr\'e model},
author = {Biddle, J. and Wang, B. and Priour, D. J. and Das Sarma, S.},
journal = {Phys. Rev. A},
volume = {80},
issue = {2},
pages = {021603},
numpages = {4},
year = {2009},
month = {Aug},
publisher = {American Physical Society},
doi = {10.1103/PhysRevA.80.021603},
url = {https://link.aps.org/doi/10.1103/PhysRevA.80.021603}
}

@article{biddle2010predicted,
title = {Predicted Mobility Edges in One-Dimensional Incommensurate Optical Lattices: An Exactly Solvable Model of Anderson Localization},
author = {Biddle, J. and Das Sarma, S.},
journal = {Phys. Rev. Lett.},
volume = {104},
issue = {7},
pages = {070601},
numpages = {4},
year = {2010},
month = {Feb},
publisher = {American Physical Society},
doi = {10.1103/PhysRevLett.104.070601},
url = {https://link.aps.org/doi/10.1103/PhysRevLett.104.070601}
}

@article{biddle2011localizationB,
title = {Localization in one-dimensional lattices with non-nearest-neighbor hopping: Generalized Anderson and Aubry-Andr\'e models},
author = {Biddle, J. and Priour, D. J. and Wang, B. and Das Sarma, S.},
journal = {Phys. Rev. B},
volume = {83},
issue = {7},
pages = {075105},
numpages = {22},
year = {2011},
month = {Feb},
publisher = {American Physical Society},
doi = {10.1103/PhysRevB.83.075105},
url = {https://link.aps.org/doi/10.1103/PhysRevB.83.075105}
}

@article{li2017mobility,
title = {Mobility edges in one-dimensional bichromatic incommensurate potentials},
author = {Li, Xiao and Li, Xiaopeng and Das Sarma, S.},
journal = {Phys. Rev. B},
volume = {96},
issue = {8},
pages = {085119},
numpages = {18},
year = {2017},
month = {Aug},
publisher = {American Physical Society},
doi = {10.1103/PhysRevB.96.085119},
url = {https://link.aps.org/doi/10.1103/PhysRevB.96.085119}
}

@article{li2020mobility,
title = {Mobility edge and intermediate phase in one-dimensional incommensurate lattice potentials},
author = {Li, Xiao and Das Sarma, S.},
journal = {Phys. Rev. B},
volume = {101},
issue = {6},
pages = {064203},
numpages = {16},
year = {2020},
month = {Feb},
publisher = {American Physical Society},
doi = {10.1103/PhysRevB.101.064203},
url = {https://link.aps.org/doi/10.1103/PhysRevB.101.064203}
}

@article{wilkinson1994spectral,
title = {Spectral dimension and dynamics for Harper's equation},
author = {Wilkinson, Michael and Austin, Elizabeth J.},
journal = {Phys. Rev. B},
volume = {50},
issue = {3},
pages = {1420--1429},
numpages = {0},
year = {1994},
month = {Jul},
publisher = {American Physical Society},
doi = {10.1103/PhysRevB.50.1420},
url = {https://link.aps.org/doi/10.1103/PhysRevB.50.1420}
}

@article{8txw-m2cp,
  title = {Quasiperiodicity-induced bulk localization with self-similarity in non-Hermitian systems},
  author = {Wang, Yu-Peng and Chang, Chuo-Kai and Okugawa, Ryo and Hsu, Chen-Hsuan},
  journal = {Phys. Rev. Res.},
  volume = {7},
  issue = {4},
  pages = {043353},
  numpages = {14},
  year = {2025},
  month = {Dec},
  publisher = {American Physical Society},
  doi = {10.1103/8txw-m2cp},
  url = {https://link.aps.org/doi/10.1103/8txw-m2cp}
}

@article{lellouch2014localization,
title = {Localization transition in weakly interacting Bose superfluids in one-dimensional quasiperdiodic lattices},
author = {Lellouch, Samuel and Sanchez-Palencia, Laurent},
journal = {Phys. Rev. A},
volume = {90},
issue = {6},
pages = {061602(R)},
numpages = {5},
year = {2014},
month = {Dec},
publisher = {American Physical Society},
doi = {10.1103/PhysRevA.90.061602},
url = {https://link.aps.org/doi/10.1103/PhysRevA.90.061602}
}

@article{yao2019critical,
title = {Critical Behavior and Fractality in Shallow One-Dimensional Quasiperiodic Potentials},
author = {Yao, Hepeng and Khoudli, Alice and Bresque, L\'ea and Sanchez-Palencia, Laurent},
journal = {Phys. Rev. Lett.},
volume = {123},
issue = {7},
pages = {070405},
numpages = {6},
year = {2019},
month = {Aug},
publisher = {American Physical Society},
doi = {10.1103/PhysRevLett.123.070405},
url = {https://link.aps.org/doi/10.1103/PhysRevLett.123.070405}
}

@article{liu2022anomalous,
title={{Anomalous mobility edges in one-dimensional quasiperiodic models}},
author={Tong Liu and Xu Xia and Stefano Longhi and Laurent Sanchez-Palencia},
journal={SciPost Phys.},
volume={12},
pages={027},
year={2022},
publisher={SciPost},
doi={10.21468/SciPostPhys.12.1.027},
url={https://scipost.org/10.21468/SciPostPhys.12.1.027},
}

\end{document}